\def\year{2020}\relax
\documentclass[letterpaper]{article} 
\usepackage{aaai20}  
\usepackage{times}  
\usepackage{helvet} 
\usepackage{courier}  
\usepackage[hyphens]{url}  
\usepackage{graphicx} 
\usepackage{booktabs}
\usepackage{graphicx}  
\usepackage{color}
\usepackage{balance}

\usepackage{amsmath}
\usepackage{subfigure}

\usepackage{multirow}

\title{Political Partisanship and Anti-Science Attitudes in Online Discussions about Covid-19}

\author{A. Rao, F. Morstatter, M. Hu, E. Chen, K. Burghardt, E. Ferrara, K. Lerman \\
Information Sciences Institute\\University of Southern California}

\begin{document}

\maketitle

\begin{abstract}
The novel coronavirus pandemic continues to ravage communities across the US. Opinion surveys identified importance of political ideology in shaping perceptions of the pandemic and compliance with preventive measures.  
Here, we use social media data to study complexity of polarization. We analyze a large dataset of tweets related to the pandemic collected between January and May of 2020, and develop methods to classify the ideological alignment of  users along the moderacy (hardline vs moderate), political (liberal vs conservative) and science (anti-science vs pro-science) dimensions. While polarization along the science and political dimensions are correlated, politically moderate users are more likely to be aligned with the pro-science views, and politically hardline users with anti-science views. Contrary to expectations, we do not find that polarization grows over time; instead, we see increasing activity by moderate pro-science users. We also show that anti-science conservatives tend to tweet from the Southern US, while anti-science moderates from the Western states. Our findings shed light on the multi-dimensional nature of polarization, and the feasibility of tracking polarized opinions about the pandemic across time and space through social media data.   

\end{abstract}

\noindent 
\section{Introduction}
Effective response to a health crisis requires society to forge a consensus on many levels: scientists and doctors have to learn about the disease and quickly and accurately communicate their research findings to others; public health professionals and policy experts have to translate the research into policies and regulations for the public to follow; and people have to follow guidelines to reduce infection threat. However, the fast-moving COVID-19 pandemic has brought into sharp relief our critical vulnerabilities at all these levels. Instead of orderly consensus-building, we have seen disagreement and controversy that exacerbated the disease. Research papers are rushed through the review, with results sometimes disputed or retracted \footnote{\url{https://retractionwatch.com/retracted-coronavirus-covid-19-papers/}}; policy makers give conflicting advice; scientists and many in the public disagree on many issues---from the benefits of therapeutics to the need for lockdowns and face covering. The conflicting viewpoints create conditions for polarization to color perceptions of the pandemic. 

Polarization is characterized by division of a group by sharply contrasting opinions. While diverse opinions are necessary in a healthy society, political scientists have observed a phenomenon called \textit{pernicious polarization}, where one social or cultural ideology precludes the possibility of a rational public discussion on the topic ~\cite{mccoy2016polarized}. Surveys have identified a partisan gulf in the attitudes about COVID-19 and the costs and benefits of mitigation strategies. According to a Pew Report~\cite{pew2020partisan},  partisanship 
significantly affects perceptions of public health measures.
Polarization has colored the messages of US political leaders about the pandemic~\cite{green2020elusive}, as well as discussions of ordinary social media users~\cite{jiang2020political}. Coupled with a distrust of science and institutions, polarization can have a real human cost if it leads the public to minimize the benefits face coverings 
or reject a vaccine when it becomes available. 


Current research measures polarization as divergence of opinions along the political dimension and its effect on other opinions, for example, discussion of scientific topics~\cite{bessi2016users}. However, opinions on controversial issues are often correlated \cite{baumann2020modeling}: for example, those who support transgender rights also believe in marriage equality, and those who oppose lockdowns also resist universal face covering. Inspired by this idea, we capture some of the complexity of polarization by projecting opinions in a multi-dimensional space, with different axes corresponding to different semantic dimensions. Once we identify the dimensions of polarization and define how to measure them, we can study dynamics of polarization and interactions between opinions.

Our work analyzes tweets related to the COVID-19 pandemic collected between January 21, 2020 and May 01, 2020~\cite{chen2020tracking}. We study polarization along three dimensions---  political (liberal vs conservative) and science (pro-science vs anti-science), and moderacy (hardline vs moderate). User alignments along the science axis identifies the polarization over scientific discussions related to the pandemic. A user's political identity is defined in a two dimensional space. 
Working in tandem with the political axis, the moderacy dimension recognizes the intensity of political standing from hardline to moderate. For the hardliners identified along moderacy dimension we leverage the political axis to identify their partisanship as Liberal or Conservative.

Leveraging a set of media sources that have been classified by nonpartisan sites along these dimensions, we identify a \textit{seed-set} 
that define the poles of  each dimension. These media sources include mainstream news and a large variety of other sources, such as government agencies,  non-governmental organizations, crowdsourced content, and alternative medicine news and health sites. We then describe network and content based inference methods to classify users along these multiple dimensions of polarization. 
%
Inferring the polarization of users discussing COVID-19 allows us to study the relationships between polarized ideologies and their temporal and geographic distributions. We show that political and science dimensions are highly correlated and that politically hardline users are more likely to be anti-science, while politically moderate users are more often pro-science. We also identify regions of the US and timepoints where the different ideological subgroups are comparably more active and their topics of conversation.


The {contributions} of this work are as follows:
\begin{itemize}
    \item We describe a framework for quantifying multi-dimensional polarization.
    \item We describe two novel methods to infer multi-dimensional polarization from the text of online conversations and compare their performance to state-of-the-art methods.
    \item We provide empirical evidence for multi-dimensional polarization in Twitter conversations about COVID-19.  
    \item We study the relationships between these dimensions, showing that political and science dimensions are highly correlated.
    \item We study the geographical distribution of users with polarized opinions and  identify US states with larger proportion of Twitter users along each polarized axis.
\end{itemize}

As the amount of COVID-19 information explodes, we need the ability to proactively identify emerging areas of polarization and controversy. Early identification will lead to  more effective interventions to reduce polarization and also improve the efficacy of disease mitigation strategies.

\section{Related Work}
Polarization is a well-recognized issue spanning the fields of psychology \cite{Myers1976,Isenberg1986}, political science \cite{mccoy2016polarized}, and even physics \cite{Sasahara2020}, as well as the present context of computer science \cite{conover2011political}. 
The foundation of polarization starts with initial studies in psychology on group polarization \cite{Myers1976,Isenberg1986}, in which opinions of a group become more extreme than initial opinions of each individual. In the present context, this could help explain why initially moderate individuals become more entrenched in the left, right, or anti-science domains.
Polarization is also explored in political science in order to explain its potential effects on government efficiency and democracy \cite{mccoy2016polarized,somer2019transformations}. These results show commonly negative effects of polarization on governments, thus motivating many explorations into this field. Moreover, they distinguish political polarization---the effect of polarization on elections---from societal polarization---its effect on social connections, with extreme polarization affecting both.

While polarization has traditionally been measured using surveys \cite{pew2020partisan}, in recent years researchers have instead begun to measure polarization with social media discussions \cite{conover2011political,bessi2016users,Schmidt2017,Bail2018}. Three consistent effects in social media have been observed. First, there is strong polarization in what people consume, measured directly from content or indirectly from retweets \cite{conover2011political,smith2013role,Schmidt2017}. In other words, people seem to selectively confine what they watch due to, e.g., confirmation bias \cite{Nickerson1998}, thus exacerbating polarization. Moreover, viewing information from the other ideological side does not necessarily affect ones opinion \cite{Bail2018}. Second, polarization is seen in different fields not directly related to left-right political polarization. This includes climate science \cite{tyagi2020affective} and even opinions about the COVID-19 epidemic \cite{pew2020partisan}. Our goal in this paper is to combine these separate findings to better understand how some of these polarizations, such as pro- or anti-science and left and right polarization, relate to one another. 

Finally, polarization has had real-world consequences outside of elections, by shaping people's perceptions of the pandemic \cite{chen2020tracking,green2020elusive,jiang2020political}. Recent research finds, for example, that polarization has affected the language policy makers use \cite{green2020elusive}, which can drive policies in different directions due to a disunited front. Moreover, sentiment towards government measures \cite{jiang2020political} or towards medical professionals \cite{pew2020partisan}, has become more polarized, further reducing the efficacy of public health measures. Our work extends on these results by exploring how polarization is more generally impacting beliefs in science, and to what degree is this correlates with partisan polarization.

Sadly polarization is not altogether unexpected. Simple models of human behavior that were inspired by psychology have been created over several years. Under the simple assumptions of social influence and selectively cutting ties with ideological opposites, echo chambers form with large disparate groups of people \cite{Durrett2012,baumann2020modeling,Sasahara2020,starnini2020}. This is also alike to homophily, in which users might form ties with people who are similar \cite{McPherson2001}. Echo chambers could then drive group polarization, and therefore drive the current social media landscape.

Our model work uniquely applies both network and text analysis to infer the degree of polarization in the network. For text analysis, we apply a number of candidate methods inspired by previous research, such as Latent Dirichlet Allocation (LDA)~\cite{blei2003latent}, and hashtag bag-of-words (BOW)~\cite{conover2011political}. We, however, find the best results using text embedding. Text embedding was discovered in 2013 \cite{mikolov2013efficient}, and was rapidly extended due to its uniquely useful features associated with both words and text \cite{pennington2014glove,joulin2016bag}. The basic nature of these methods are to embed text into a vector space such that nearby vectors represent semantically similar text. Finally, to analyze networks, we apply a Label Propagation Algorithm (LPA) proposed by Raghavan et al.,~\citeyear{Raghavan2007}. This algorithm, while first applied to community detection in networks, effectively assumes homophily drives how opinions form, and therefore its performance would test to some degree how homophily drives polarization.

Our work differs from these previous methods via systematic analysis of multiple polarized dimensions in social media. We also contrast with previous work via multi-dimensional polarization of COVID-19. This provides a test of some results predicted in polarization models \cite{baumann2020modeling,starnini2020}. 

\section{Dataset and Methods}
We describe the data and methods for measuring polarization and also inferring it from text and online interactions.

\subsection{Data}

In this study, we use a public 
dataset of COVID-19 tweets~\cite{chen2020tracking}. This dataset comprises of $115M$ tweets from users across the globe, 
collected over a period of 101 days from January 21, 2020 to May 01, 2020. These tweets contain at least one of a predetermined set of COVID-19-related keywords (e.g., coronavirus, pandemic, Wuhan, etc.). We specifically focus on tweets from users located in the US (at state-level granularity) based on information contained in their profile and tweets. 
This geo-referenced dataset consists of 27M tweets posted by 2.4M users over the 101 day period.

\subsection{Measuring Polarization using Domain Scores}
We characterize individual opinions along three polarization dimensions. 
The \textit{political} dimension, standard dimension for characterizing partisan polarization, captures the difference between \textit{Left}/ \textit{liberal} and \textit{Right}/\textit{conservative} political stances for users with Hardline opinions.  The \textit{science} dimension captures an individual's acceptance of evidence-based \textit{Pro-Science} views or the propensity to hold \textit{Anti-Science} views. People believing and promoting conspiracies, especially health-related and pseudo-scientific conspiracies, are often grouped in the anti-science camp. Finally, the \textit{moderacy} dimension describes the intensity of partisanship---from \textit{Moderate} or neutral opinions to politically \textit{Hardline} opinions. 


We begin with a set of web domains of curated information sources that were labeled along these dimensions by non-partisan organizations, such as \textit{Media Bias-Fact Check} ({\url{https://mediabiasfactcheck.com/}}), Allsides ({\url{https://https://www.allsides.com/}}) and Newsguard, which tracks coronavirus misinformation. 
Table \ref{tab:Domain_Score} lists exemplar domains in each category. 
Domains listed under ``Conspiracy'' and ``Questionable Sources''  are mapped to our Anti-Science category.  For the Moderacy axis, we consider the union of Left and Right domains as a proxy for Hardline category, while union of ``Least-Biased'', ``Left-Moderate'' and ``Right-Moderate'' domains form the proxy Moderate category. 

\begin{table}[tbh]
\centering
    \footnotesize
\begin{tabular}{p{1.3cm} p{2.3cm} p{3.8cm}}\hline
  Dimension & Polarization along dimension & Domains \\
  \hline
  \multirow{8}{*}{Science} & \multirow{4}{*}{Pro-Science ($+1$)} & cdc.gov, who.int, thelancet.com, mayoclinic.org, nature.com, newscientist.com $\ldots$ (150+ domains) \hfill \\
  & \multirow{4}{*}{Anti-Science ($-1$)} & 911truth.org, althealth-works\-.com, naturalcures.com, shoe\-bat\-.com, prison-planet\-.com $\ldots$ (450+ domains) \hfill\\
  \hline
  \multirow{8}{*}{Political} & \multirow{4}{*}{Liberal $(-1)$} & 
  democracynow.org, huffington-post.com, newyorker.com, occupy.com, rawstory.com,$\ldots$ (300+ domains)\\ 
  & \multirow{4}{*}{Conservative $(+1)$} & nationalreview.com, news-max.com, oann.com, theepochtimes.com, bluelivesmatter.blue $\ldots$ (250+domains)\\
  \hline
    \multirow{8}{*}{Moderacy} & \multirow{4}{*}{Moderate $(+1)$} & ballotpedia.org, c-span.org, hbr.org, wikipedia.org, weforum.org, snopes.com, reuters\-.com $\ldots$ (400+ domains)\\ 
    &  \multirow{4}{*}{Hardline $(-1)$} & gopusa.com, cnn.com, demo\-cracy\-now\-.org, huffington-post\-.com, oann.com, the\-epoch\-times.com $\ldots$ (500+ domains)\\
  \hline
\end{tabular}
\caption{Curated information and news domains with their polarization. Pro-Science domains are mapped to $+1$ along the science axis while, Anti-Science domains are mapped to $-1$. Along the political axis, Liberal domains are mapped to $-1$ while, Conservative ones are mapped to $+1$. On the moderacy axis, we map Hardline domains as $-1$ and Moderate domains as $+1$.}
\label{tab:Domain_Score}
\end{table}

We quantify a user's position along the dimensions of polarization by tracking the number of links to curated domains the user shares.   
Specifically, we extract 
domains shared by users in the 101 day period and filter for relevant domains present in our curated list of domains  (Table \ref{tab:Domain_Score}). This gives us a set of 136K users who shared Science domains, 
169K
users who shared Political domains and 234K 
users who shared domains along the Moderacy dimension. 
After filtering out users who shared fewer than \textit{three} relevant domains, this leaves us with 18.7K 
users who have shared domains across all three dimensions. For each user, we compute a \textit{domain score} $\delta$ along each of the three dimensions, as the average of mapped domain values of a dimension:
$$\delta_i = \frac{\Sigma D_{i,d}}{|D_{i,d}|};\,\, \forall d \in \{\texttt{Science},\,\texttt{Political},\,\texttt{Moderacy}\}
$$
where, $\delta_i$ is the domain score of $user_{i}$ and $D_{i,d}$ represents the set of domains shared by $user_{i}$ relevant to dimension $d$.

\begin{figure*}[tb]
\centering
    \includegraphics[width=0.9\linewidth]{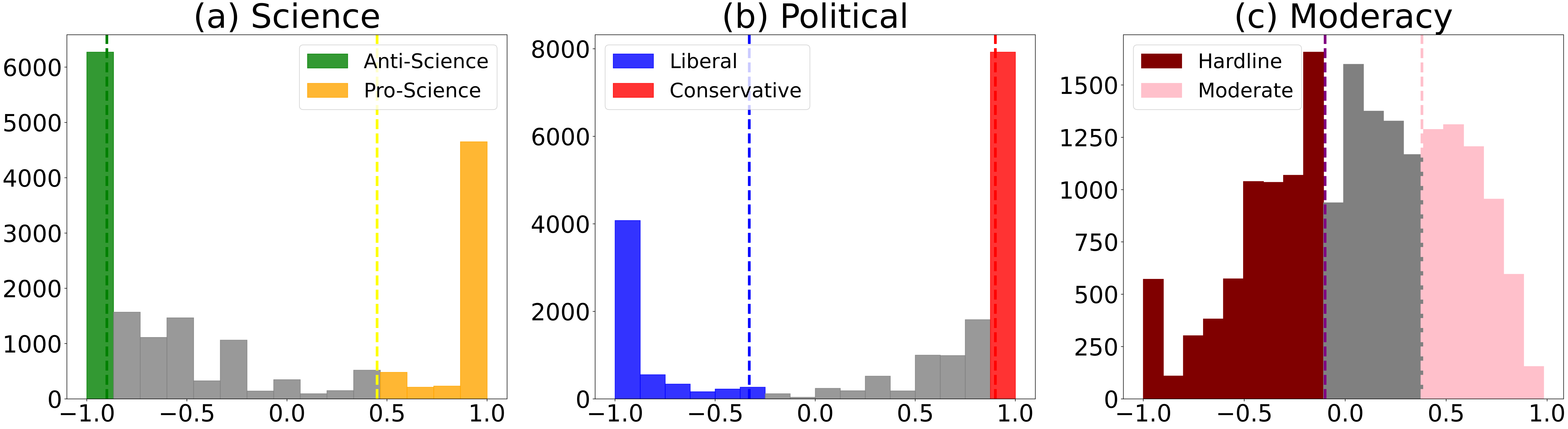}
    \caption{The distribution of domain scores along Science, Political and Moderacy dimensions. The vertical lines at $0.42$ and $-1$ in (a), marks the top and bottom $30\%$ cutoffs of distribution along the Science dimension which are binned as $Pro-Science (+1)$ and $Anti-Science(-1)$, respectively. The vertical lines at $1$ and $-0.33$ in (b), marks the top and bottom $30\%$ cutoffs of distribution along the Political dimension which are binned as $Conservative(+1)$ and $Liberal(-1)$, respectively. The vertical lines in (c) at $0.38$ and $-0.18$ indicates the top and bottom $30\%$ cutoffs of distribution along Moderacy dimension which are binned as $Moderate (+1)$ and $Hardline(-1)$, respectively.}
    \label{fig:Domain-Score}
\end{figure*} 

Figure~\ref{fig:Domain-Score} shows the distribution of domain scores across the three dimensions for users who post domains across all three dimensions of interest. The distribution are peaked at their extreme values, showing more users sharing information from Anti-Science than Pro-Science domains  and more Conservative than Liberal domains. 


For network level analysis, we then built a web scraper that maps domains to their respective Twitter handles. 
The scraper initiates a simple Google query of the form ``\textit{$Domain\,\,Name$ Twitter Handle}". This tool relies on the search engine to rank results based on relevance and picks out the title of the first result containing the sub-string ``$ | Twitter$". This substring is of the form ``\textit{$Account\,\,Name\,\,(@handle) | $ Twitter}" which is parsed to retrieve the domain's corresponding handle. We manually verified the mapped domains.

The mapped Twitter handles form our seed sets for semi-supervised learning at the network level. Each dimension's seed-set comprises key-value pairs of Twitter handles and their corresponding orientation along the dimension. Table \ref{tab:LPA_Seeds} illustrates the number of seeds along each polarization axis.

\subsection{Inferring Polarization}
\label{sec:inferring_polarization}
Using domain scores, we can quantify the polarization of just a small fraction (0.7\%) of users in the dataset. In this section we describe how we leverage this data to infer the polarization of the remaining users in our dataset along multiple dimensions. In the results, we compare the performance of these inference methods. 

We classify binned domain scores along each dimension because we find classifiers work better than regression in this dataset
. In light of this fact, we bin the extreme ends of domain score distribution into two classes along each dimension as shown in Figure \ref{fig:Domain-Score}.



\subsubsection{Label Propagation}
\label{sec:lpa}
Label propagation was used in the past to label user ideology based on the ideology of accounts the user retweets (see, for example \cite{badawy2018analyzing}). The idea behind label propagation is that people prefer to connect to---and retweet content posted by---others who share their opinions
~\cite{boyd2010tweet,metaxas2015retweets}. 
This gives us an opportunity to leverage topological information from the retweet network to infer users propensity to orient themselves along ideological dimensions.

\begin{table}
\centering
    \footnotesize
\begin{tabular}{lll|ll}
\toprule
  \textbf{Dim} & \textbf{Polarization} & \textbf{Seeds} & \textbf{Statistic} & \textbf{Value} \\
  \midrule
  \multirow{2}{*}{Science} & \multirow{1}{*}{Pro-Science} & $81$ & Nodes  & 1,857,028\\ \cline{2-5}
  & \multirow{1}{*}{Anti-Science} & $77$ & In-deg & 39,149\\
  \hline
  \multirow{2}{*}{Political} & \multirow{1}{*}{Liberal} & $96$ & Out-deg & 1,450\\ \cline{2-5}
  & \multirow{1}{*}{Conservative} & $99$ & RTs  & 9,788,251\\
  \hline
  \multirow{2}{*}{Moderacy} &  \multirow{1}{*}{Hardline} & $195$ & Uniq RTs & 7,745,533\\ \cline{2-5}
  & \multirow{1}{*}{Moderate } & $363$ & SCC & 1,818,657\\
  \hline
\end{tabular}
\caption{Description of the retweet network. Number of seed handles along each polarization axis for initial node assignment in the LPA. Statistics of the network, including maximum in- and out-degree and size of the strongly connected component (SCC).}
\label{tab:LPA_Seeds}
\end{table}


To this end, we build a network from 9.8M 
retweet interactions between 1.9M 
users from the geocoded Twitter dataset. In the retweet network, an edge runs from $A$ to $B$ if, user $A$ retweets user $B$. Descriptive statistics of the retweet network are shown in Table \ref{tab:LPA_Seeds}. We then use a semi-supervised greedy learning algorithm to identify clustsers in the retweet network.

LPA proposed by ~\cite{Raghavan2007} is a widely-used near-linear time network community detection algorithm. This greedy learning approach starts off with an initial random label assignment and, iteratively re-assigns labels to nodes along a dimension. This reassignment eventually converges to a state of equilibrium where all nodes in the network are assigned labels which are shared by the majority of their corresponding neighbors. However, owing to arbitrary tie-breaking, a certain amount of randomness creeps into the results produced by this algorithm. Courtesy of this stochasticity, LPA tends to generate different cluster assignments (user polarization) for the same network.

\begin{table*}
\footnotesize
\centering
\begin{tabular}{lllllll}\hline
  \toprule
  \textbf{Method} & \textbf{Dimension} & \textbf{Dataset Size}& \textbf{Accuracy}& \textbf{Precision}& \textbf{Recall}& \textbf{F1-Score} \\
  \midrule
  
  \multirow{4}{*} & \multirow{1}{*}{Science} &$158$ & $92.6\%$  &$100\%$  & $80\%$ & $88.9\%$ \\
  
  {LPA} & \multirow{1}{*}{Political} & $195$ & $92.3\%$  & $86.9\%$ & $100\%$ & $93.0\%$ \\
  
  & \multirow{1}{*}{Moderacy} & $1205$ & $20.1\%$   & $72\%$ & $1.4\%$ & $2.74\%$ \\
  
  \midrule
  
  \multirow{4}{*} & \multirow{1}{*}{Science} &$1960$ &  $88.2\%$  &$88.9\%$  & $98.8\%$ & $93.6\%$ \\
  
  {Hashtag BOW} & \multirow{1}{*}{Political} & $1610$& $72.9\%$ & $75.4\%$ & $61.1\%$ & $67.4\%$ \\
  
  & \multirow{1}{*}{Moderacy} & $4684$ & $75.1\%$  & $77.7\%$ & $90.6\%$ & $83.6\%$ \\
  
  \midrule
  
  \multirow{4}{*} & \multirow{1}{*}{Science} &$9983$ & $92.2\%$  &$91.6\%$  & $92.4\%$ & $91.9\%$ \\
  
  {LDA} & \multirow{1}{*}{Political} & $11020$ & $93.5\%$  & $95.1\%$ & $93.3\%$ & $94.2\%$ \\
  
  & \multirow{1}{*}{Moderacy} & $9565$ & $86.4\%$ & $85.6\%$ & $85.0\%$ & $85.4\%$ \\
  
  \midrule
  
  \multirow{4}{*} & \multirow{1}{*}{Science} &$11202$& {$\textbf{93.8\%}$}  & {$93.9\%$}  & {$93.7\%$} & {$\textbf{93.8\%}$} \\
  
  {\textbf{fastText}} & \multirow{1}{*}{Political} & $12425$ & $\textbf{95.1\%}$ & $96.5\%$ & $94.6\%$ & $\textbf{95.5\%}$ \\
  
  & \multirow{1}{*}{Moderacy} & $11197$ & $\textbf{90.2\%}$  & $90.1\%$ & $90.5\%$ & $\textbf{90.2\%}$ \\
  
  \bottomrule
\end{tabular}
\caption{{Performance of Polarization Classification. Results compare classification performance of LPA and content-based methods including Hashtag BOW, topic modeling (LDA) and full text embedding (fastText). Results are averages of 5-fold cross validation.}}
\label{tab:dimaccs}
\end{table*}


\paragraph{Hashtag Bag of Words}~
\label{sec:bow}
We extend the method used by \cite{conover2011political} to measure user's political ideology to additional dimensions of interest. First, we identify seed hashtags defining the extremes of each dimension of polarization and generate \textit{vectors} of 100 hashtags that co-occur with these seeds. For the Science dimension, the two seed hashtags we picked are \textit{\#stayhome} and \textit{\#plandemic}. For the Political dimension, we identify hashtags that co-occur with  the seeds \textit{\#trumpvirus} and  \textit{\#chinavirus}; and for the Moderacy dimension, we find hashtags co-occurring with two moderate seeds - \textit{\#pandemic}, \textit{\#lockdown} and two hardline seeds - \textit{\#trumpvirus}, \textit{\#chinavirus} . We then, compute the dimension-wise TF-IDF transformations of corresponding vectors to generate feature vectors for each user, representative of each user's content along a dimension over time.






However, this method suffers from a critical shortcoming: a significant number of users don't use any of the hashtags in our seed sets. This results in empty feature vectors. Considering only users with non-zero feature vectors and domain scores along each dimension gives us 1.96K 
users along the Science dimension, 1.6K
users along the Political dimension and 4.6K 
users along the Moderacy dimension. 

To infer user polarization, we train a simple Logistic Regression model on the feature vector matrix with the domain scores as ground truth. 

\paragraph{Latent Dirichlet Allocation}
To reduce the dimensionality of hashtag feature vectors, we use LDA~\cite{blei2003latent} to identify topics, or groups of hashtags, and represent users as vectors in this topic space. 
In contrast to the hashtag BOW method, we consider \emph{all} hashtags generated by a user as a document representing that user (after ignoring hashtags used by fewer than $10$ users or more than $75\%$ of the users)---leaving us with 25.2K hashtags. 
We use 20 topics, as that gives the higher coherence score.



We use the document-topic affinity matrix generated by LDA to represent users. An \textit{affinity vector} is comprised of $20$ likelihood scores, adding up to 1, with each score indicating the probability of corresponding topic being a suitable representation for the set of hashtags generated by the user.
Using these affinity vectors, we generate feature vector matrices for each of the three dimensions of interest. In doing so, we can represent over 900K users who use some hashtag in their tweets with a dense vector of length $20$. 

\subsubsection{Text Embedding}
Previous methods (see \cite{conover2011political}) classified user's political polarization based on the text of their tweets by generating TF-IDF weighted unigram vectors for each user. However, the advent of more powerful text-embedding techniques~\cite{mikolov2013efficient,pennington2014glove,joulin2016bag} allows us to generate sentence embedding vectors to better represent content.

We group the tweets generated by each of the 2.4M users over a 101-day period from January to May 2020. More specifically, we collect all tweets generated by a user in this time period and concatenate them to form a text document for each user. After preprocessing the 2.4M documents to remove hashtags, URLs, mentions, handles and stopwords, we use fastText sentence embedding model pretrained on Twitter data, to generate tweet embeddings for each user. The \textit{Sent2vec} Python package \cite{DBLP:conf/naacl/GuptaPJ19} provides us with a Python interface to quickly leverage the pretrained model and generate 700-dimension feature vectors representing each user's discourse.

\begin{figure*}[tb]
    \centering
    \subfigure[Science vs Political dimensions]{\includegraphics[width=0.4\linewidth]{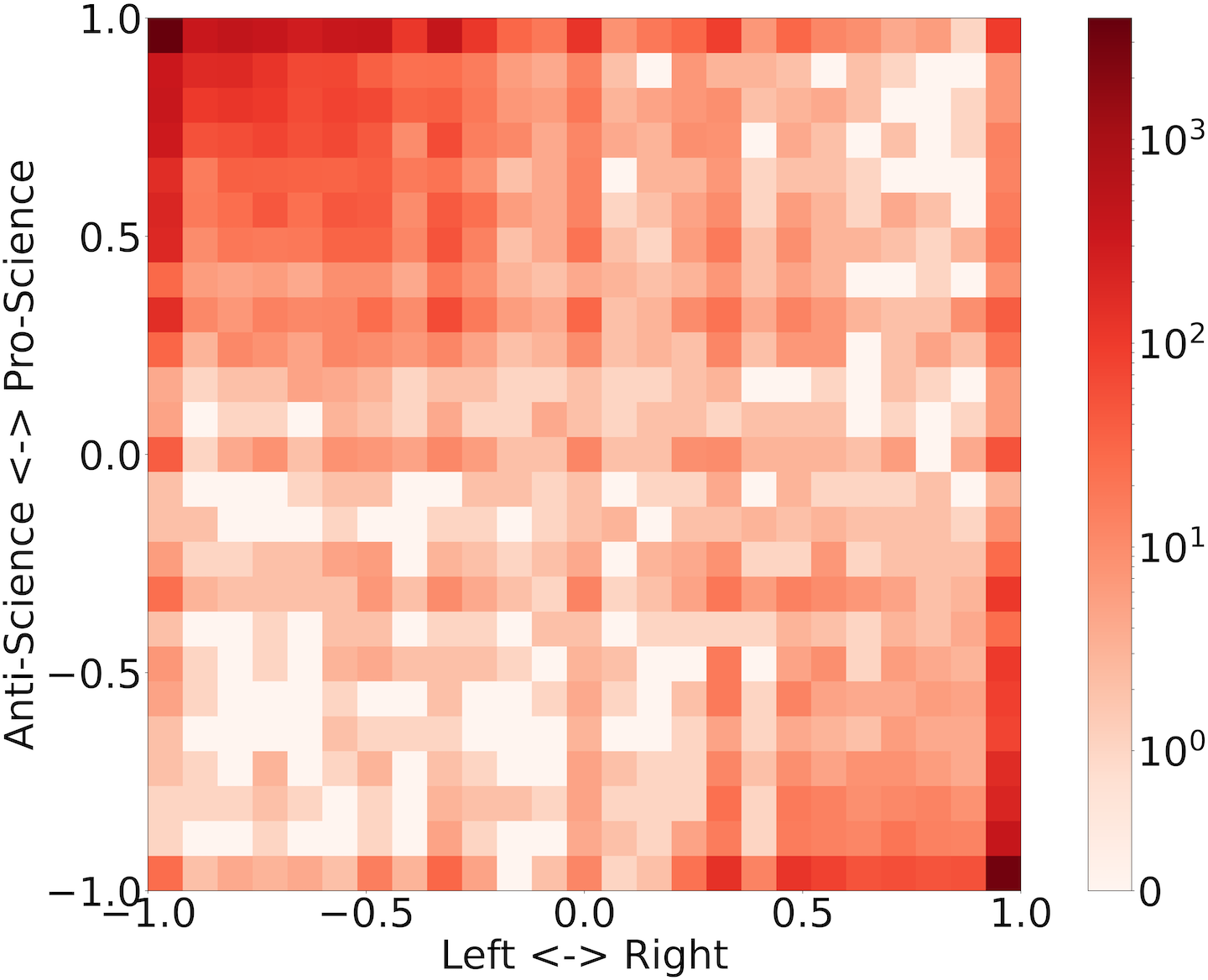}}
    \subfigure[Science vs Moderacy dimensions]{\includegraphics[width=0.4\linewidth]{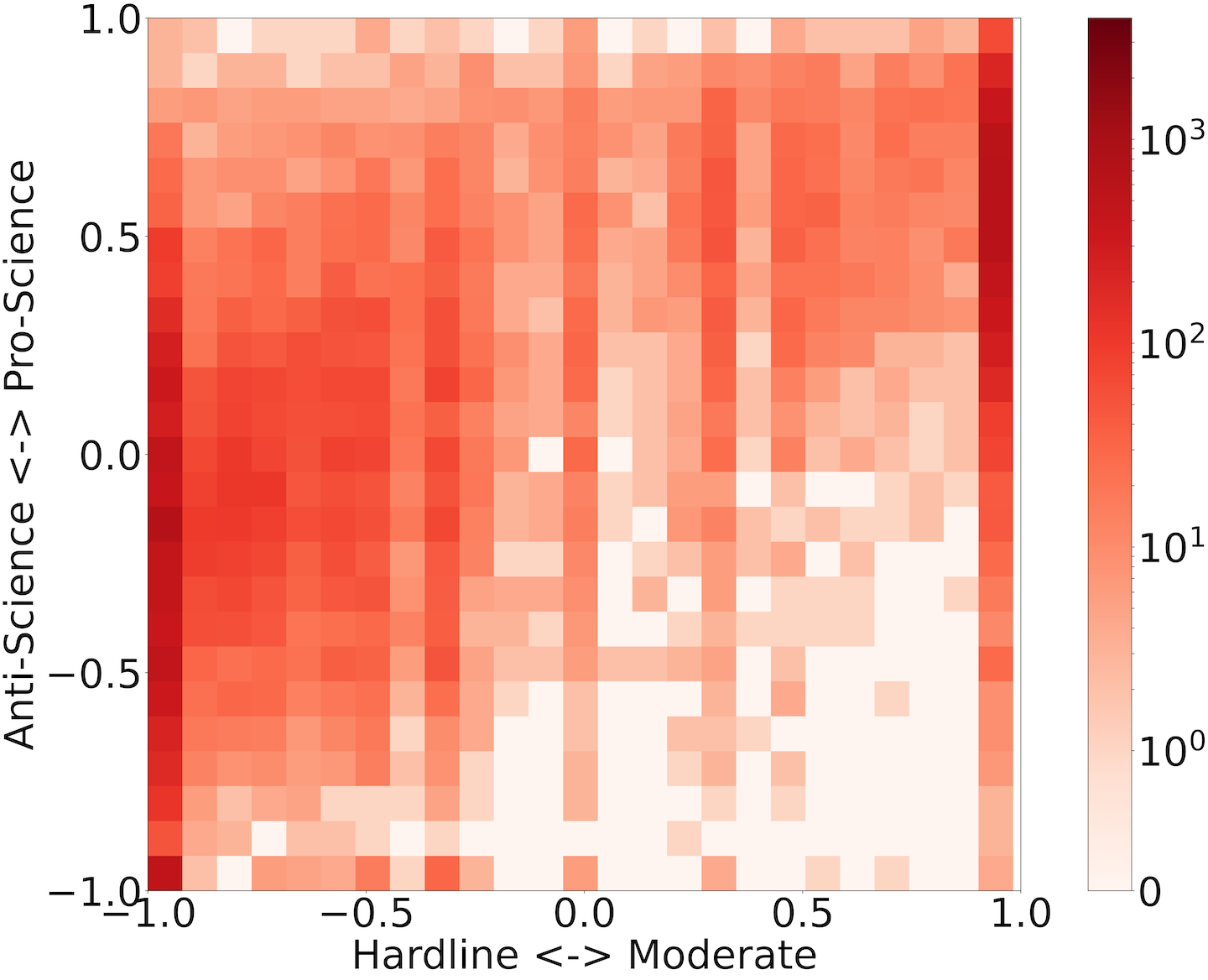}}
    \caption{Polarization of COVID-19 tweets. (a) Heatmap of polarization (domain scores) along the Science--Partisanship dimensions. Each bin within the heatmap represents the number of users with domain scores falling within that bin. (b) Heatmap of polarization along the Science--Moderacy dimensions. }
    \label{fig:heatmap}
\end{figure*}

\section{Results}

First, we visualize the domain scores of the 18.7K user, showing the relationship between the Science, Moderacy and Political dimensions. Then we compare the performance of algorithms for classifying users along the three dimensions of polarization, using domain scores as ground truth data. We used the inferred scores to study the dynamics and spatial distribution of polarized opinions of users engaged in online discussions about COVID-19.

\subsection{Visualizing Polarization}
Figure \ref{fig:heatmap} shows the relationship between dimensions of polarization, leveraging domain scores of 18.7K users who shared information from curated domains. 
The heatmap shows the density of users with specific domain scores. Large numbers of users are aligned with Pro-Science-Left (top-left corner) or Anti-Science-Right (bottom-right corner) extremes, with lower densities along the diagonal between these extremes (Fig.~\ref{fig:heatmap}(a)). This illustrates the strong correlation between political partisanship and scientific polarization, thereby highlighting the influence of pernicious political divisions on evidence-based discourse during the pandemic, with Conservatives being more likely to share Anti-Science information than Pro-Science sources. Figure \ref{fig:heatmap}(b), highlights the interplay between the Science and Moderacy axes. The white region in the bottom right corner shows there are few Anti-Science users who are politically moderate. The shading also highlights a higher density of Pro-Science users identifying as politically moderate. 

\subsection{Classifying Polarization}
To run the LPA, we start from a set of labeled seeds---Twitter handles corresponding to domains categorized along the dimensions of interest (Tables \ref{tab:Domain_Score} \& \ref{tab:LPA_Seeds}).
We reserve some of the seeds along each dimension for testing LPA predictions and report accuracy of $5$-fold cross validation. 

For content-based approaches, we used binned domain scores of 18.7K users as ground truth data to train Logistic Regression models to classify user polarization along the three dimensions. We represented each user as a vector of features generated by different content-based approaches:  hashtag frequencies for the BOW approach, topic vectors for LDA and sentence embeddings for the fastText approach. We reserved a subset of users for testing performance. 

Table \ref{tab:dimaccs} compares the performance of polarization classification methods. LPA works best when it tries to identify user alignment along the Political and Science dimensions. However, it fails to capture the subtler distinctions along the Moderacy axis. Training is further hampered by the low number of retweet interactions with Moderate domains in comparison to Hardline ones. Of the 1.8M retweet interactions, only 250K involve some Moderate seed nodes, whereas over 1M interactions involve some Hardline seed nodes. Moreover, poor classification performance with LPA reveals an important pattern: that moderates surround themselves with diverse opinion and thus a clear distinction cannot be made by observing who they retweet.  

Hashtag BOW describes users as vectors of weighted frequencies of hashtags they use. This method critically suffers from the curse of dimensionality and the lack of uniformity of usage among users. Since the dimensionality of hashtag vectors is the entire vocabulary, each vector describing a user is sparse, resulting in non-competitive performance. 

LDA modeling on hashtags allows us to generate reduced-dimension, dense feature vectors for over 900K users who use hashtags in their tweets. This representation allows us to design better learning models that significantly outperform the Hashtag BOW and LPA models.

A Logistic Regression model trained on user-text embeddings and domain scores (fastText) outperforms all other models described in this study. Coupled with fastText's ability to better handle out-of-vocabulary terms, the model's access to finer levels of detail at tweet text, culminates in it better predicting dimensions of polarization. Given the model's superior performance across all three dimensions, we leverage its predictions in subsequent analyses.

We classify users along the three polarization dimensions. However, since the definition of the Hardline extreme of the Moderacy dimension overlaps with the Political dimension, we need to report only six ideological groups, rather than all eight combinations. 

\begin{figure}[tbph]
    \centering
    \includegraphics[width=0.9\linewidth]{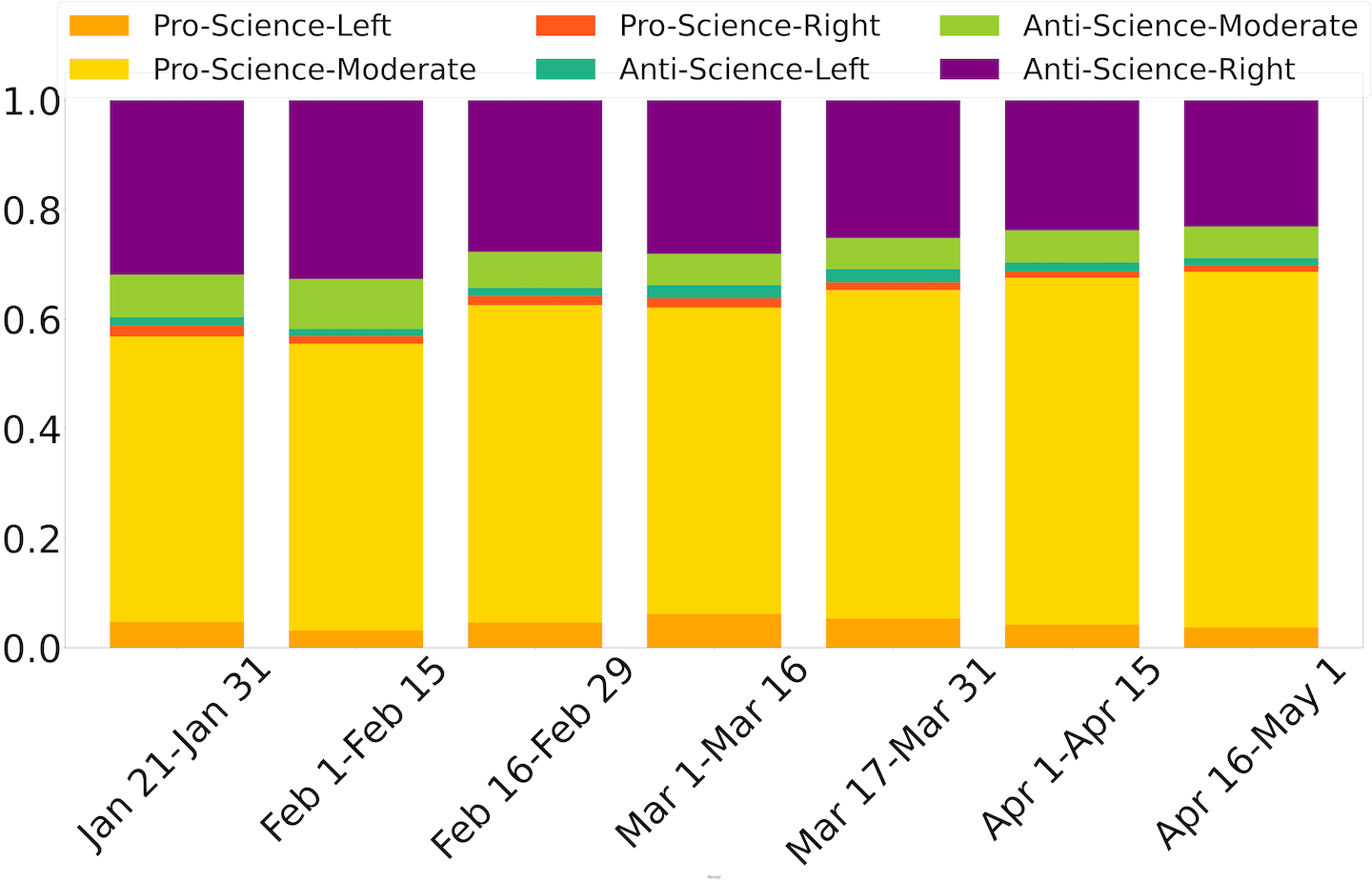}
    \caption{Fraction of active users per ideological group in bi-weekly periods.}
    \label{fig:pol-over-time}
\end{figure}

\begin{figure*}[tbh]
    \centering
    \subfigure[Pro-Science-Left]
    {\includegraphics[width=0.25\textwidth]{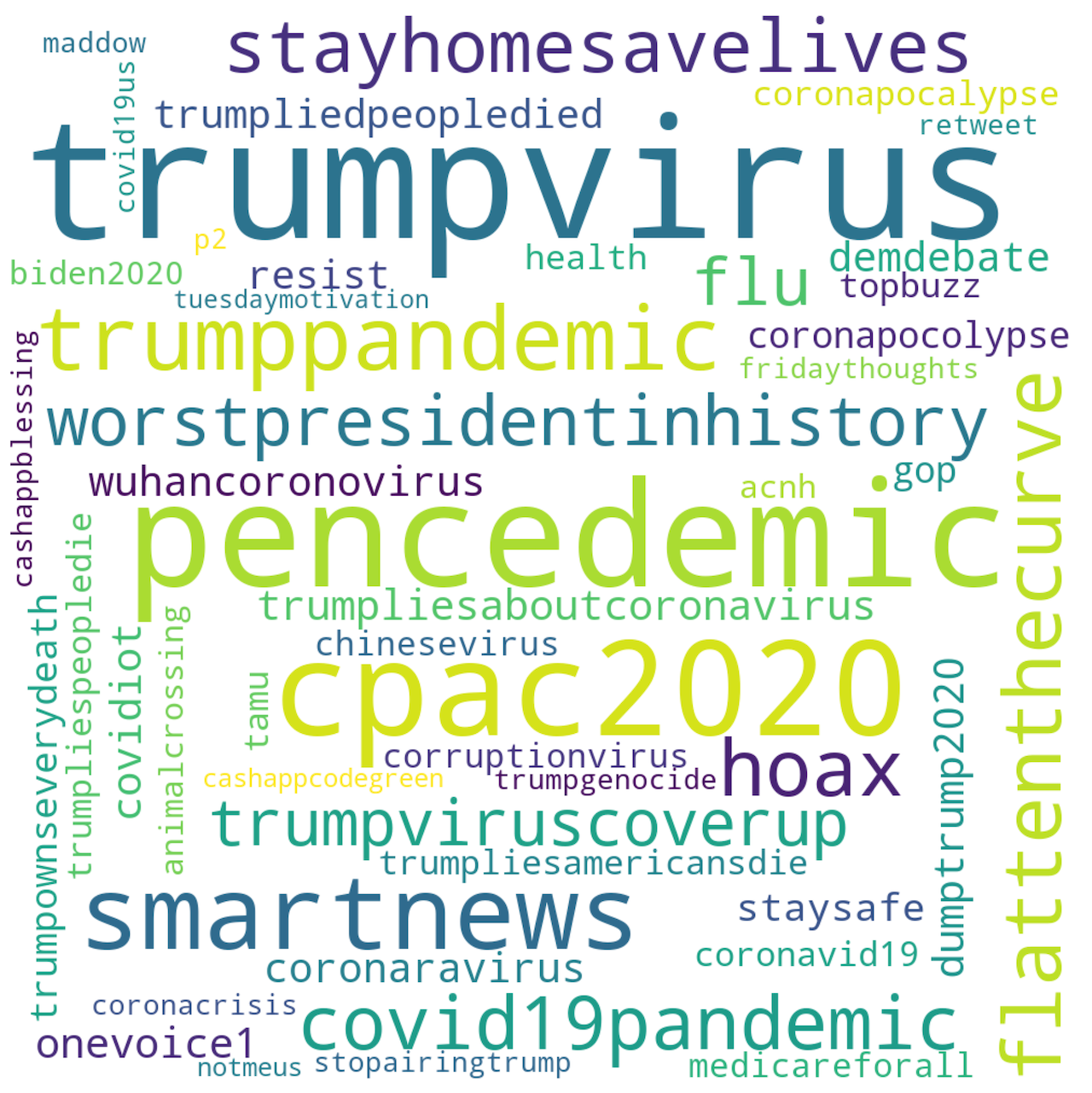}} 
    \subfigure[Pro-Science-Moderate]
    {\includegraphics[width=0.25\textwidth]{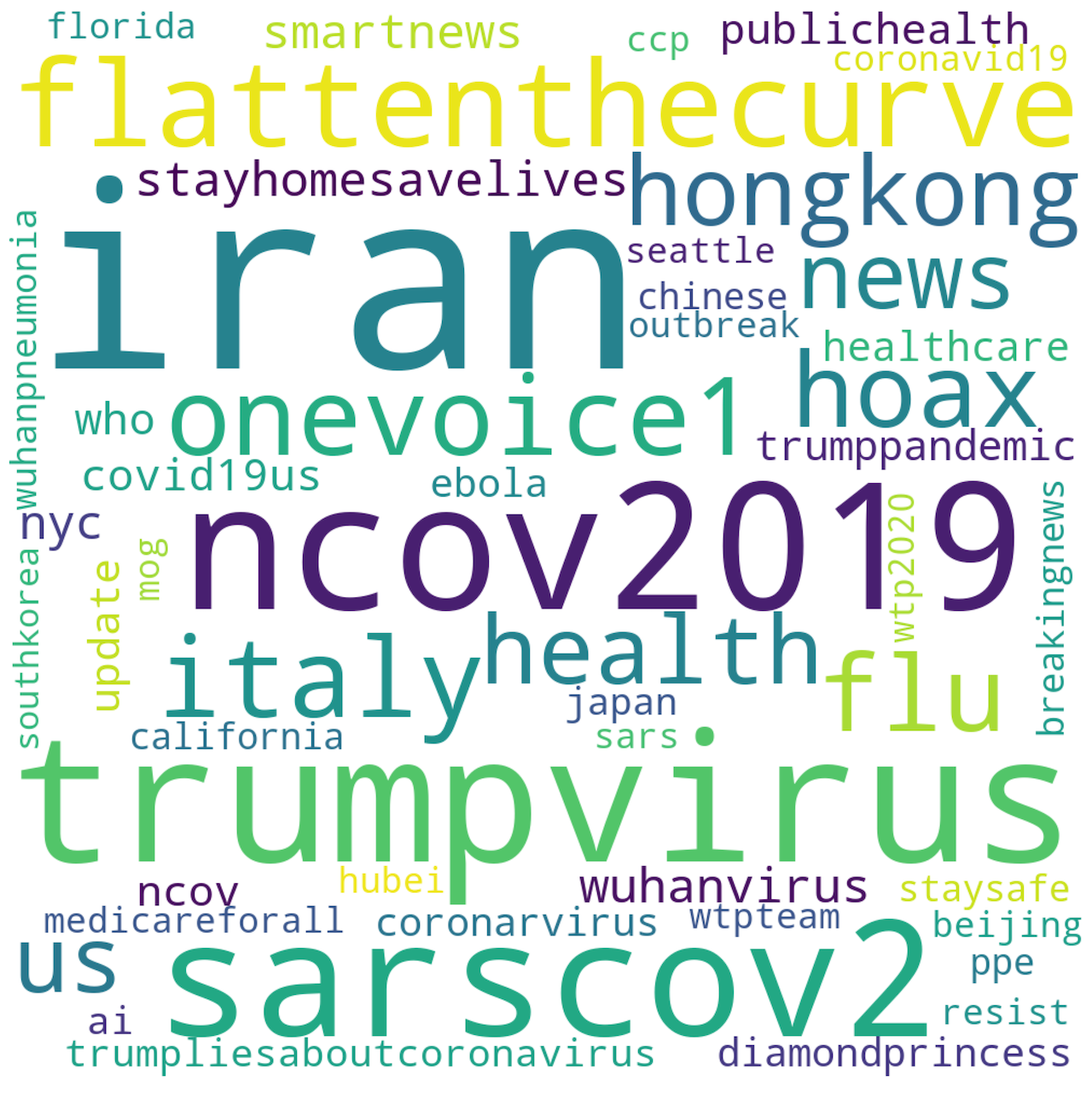}} 
    \subfigure[Pro-Science-Right]
    {\includegraphics[width=0.25\textwidth]{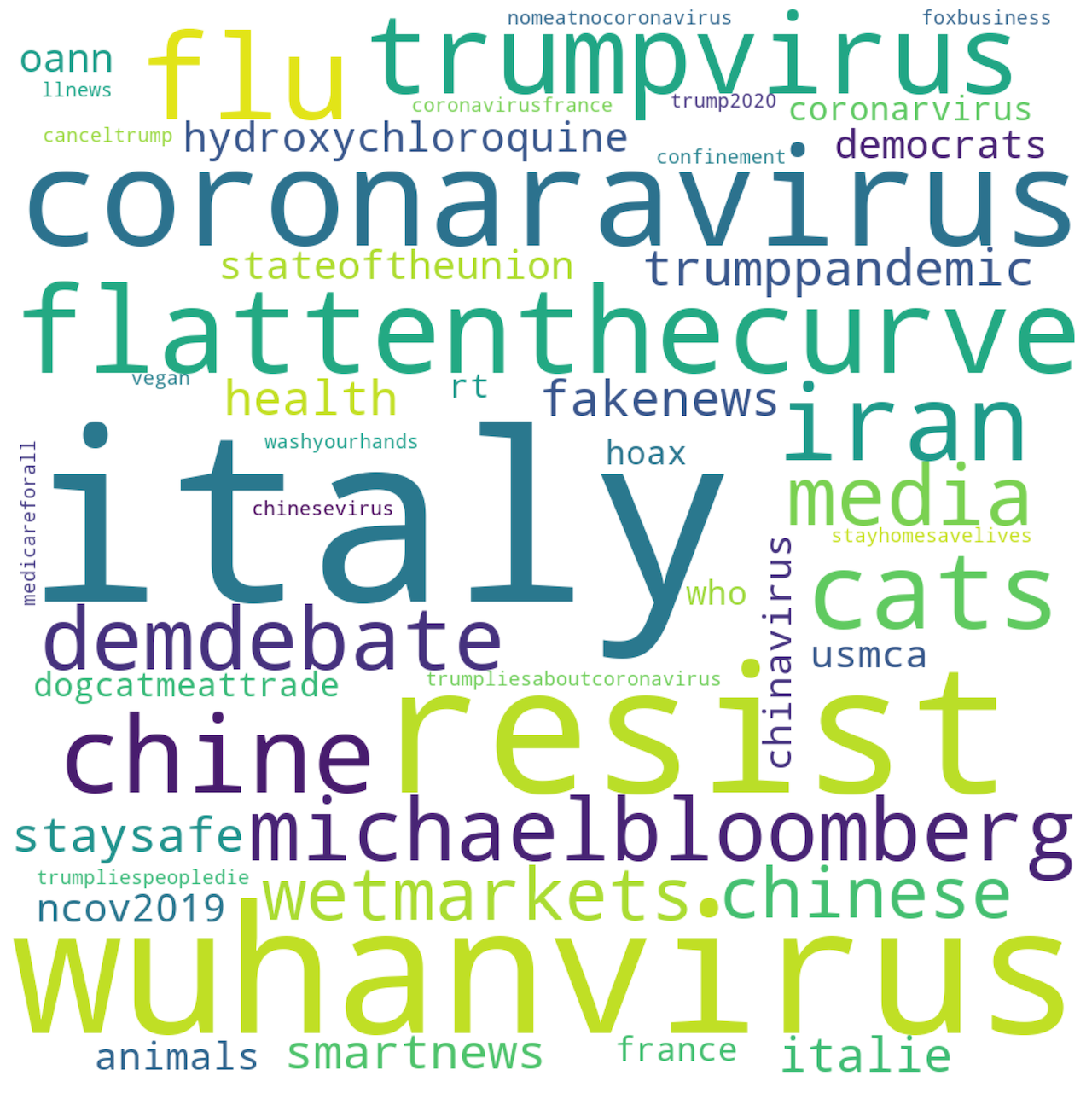}}\\
    \subfigure[Anti-Science-Left]
    {\includegraphics[width=0.25\textwidth]{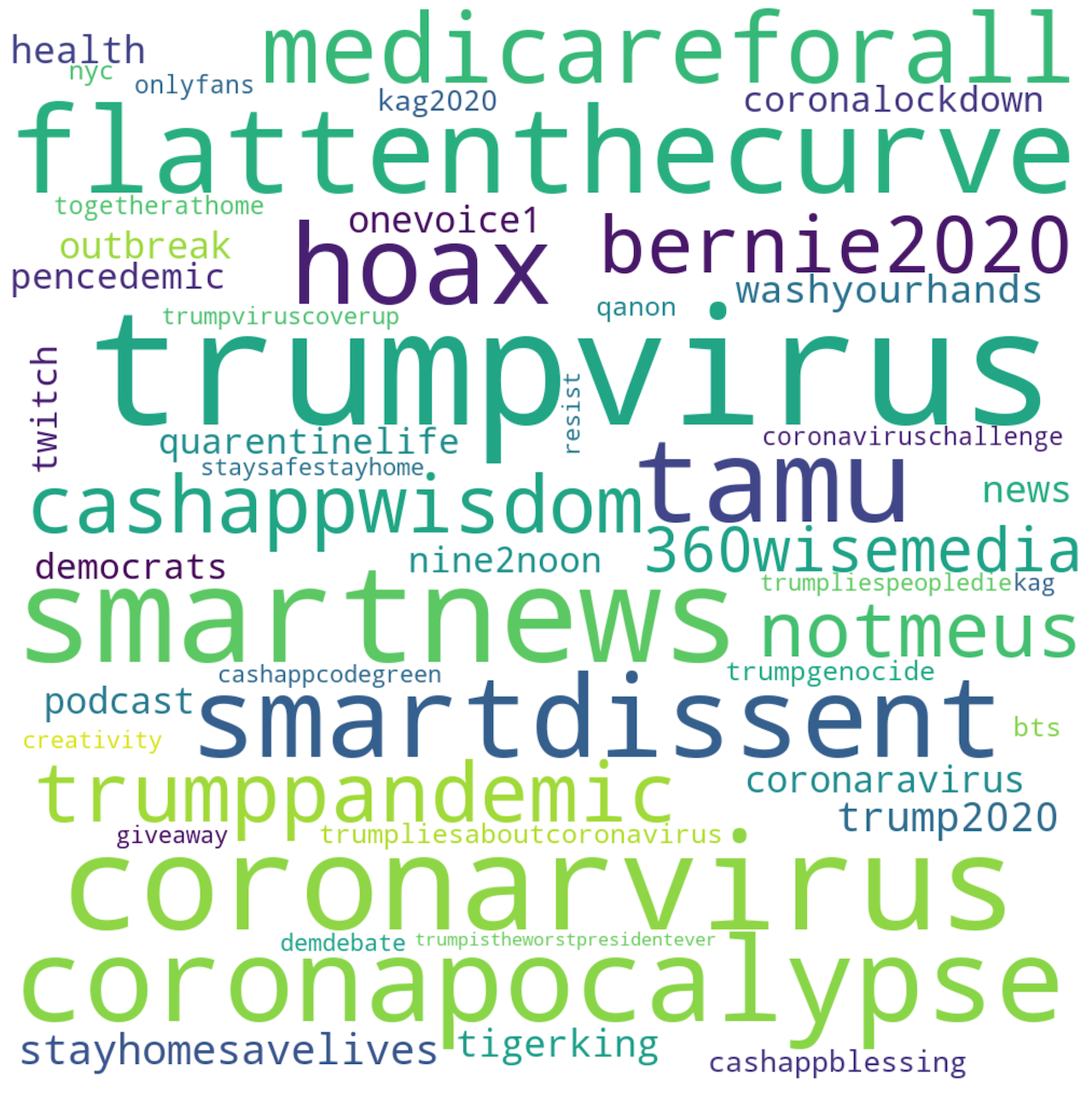}}
    \subfigure[Anti-Science-Moderate]
    {\includegraphics[width=0.25\textwidth]{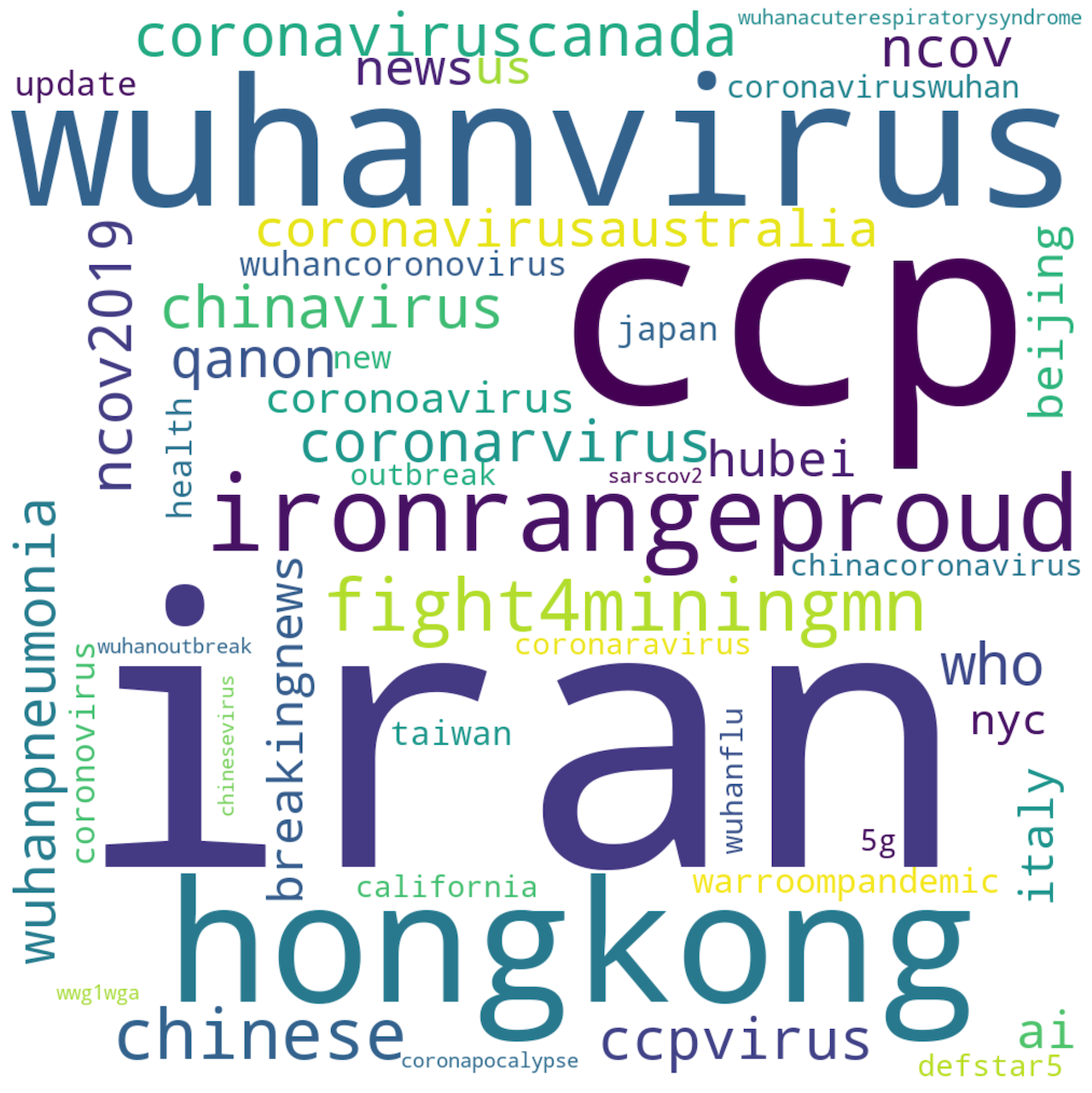}}
    \subfigure[Anti-Science-Right]
    {\includegraphics[width=0.25\textwidth]{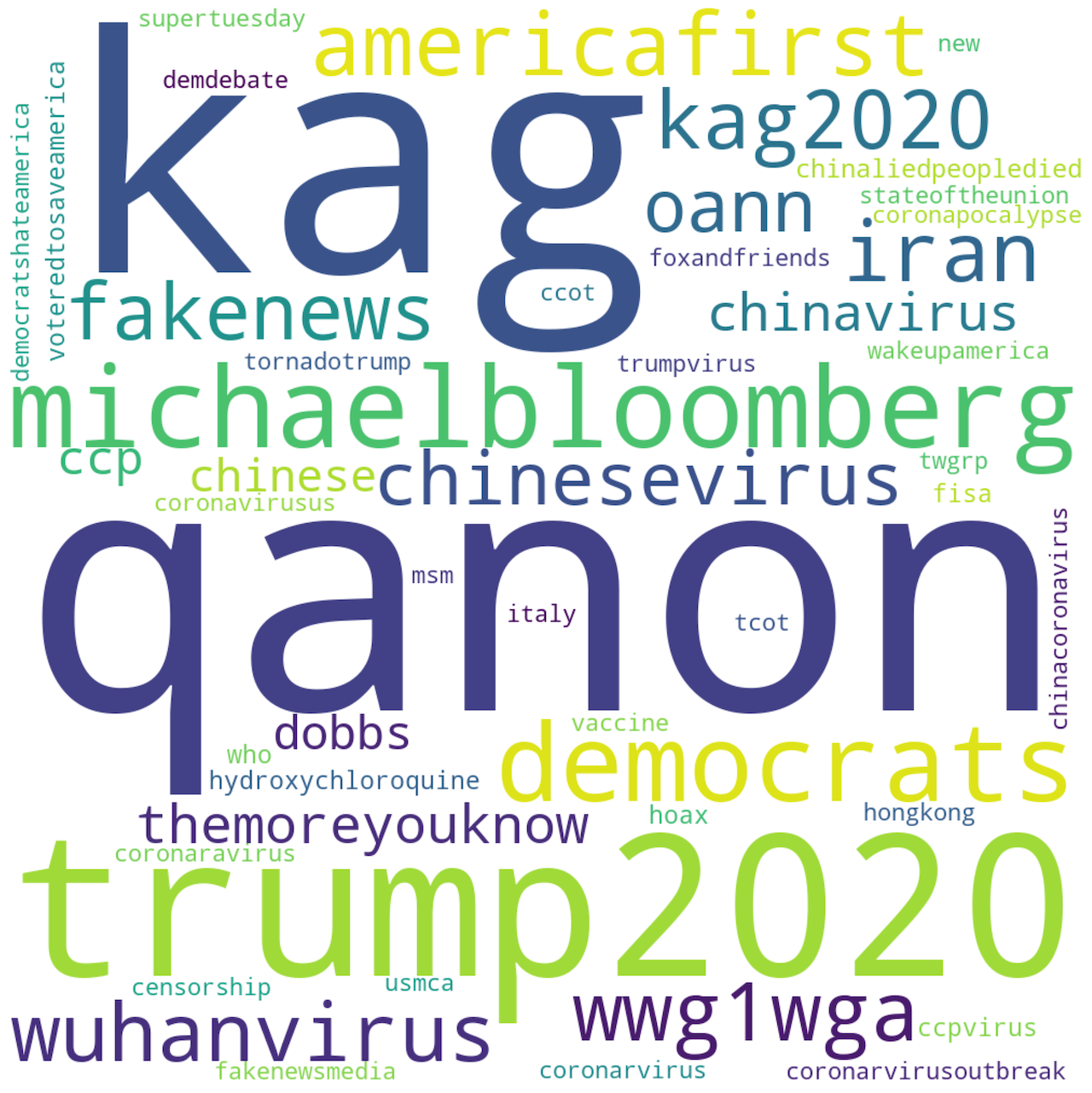}}
    \caption{Topics of discussion within the six ideological groups.}
    \label{fig:hashtag-wordcloud}
\end{figure*}
\begin{figure*}[tbh!]
    \centering
    \subfigure[Pro-Science-Left ]{\includegraphics[width=0.3\textwidth]{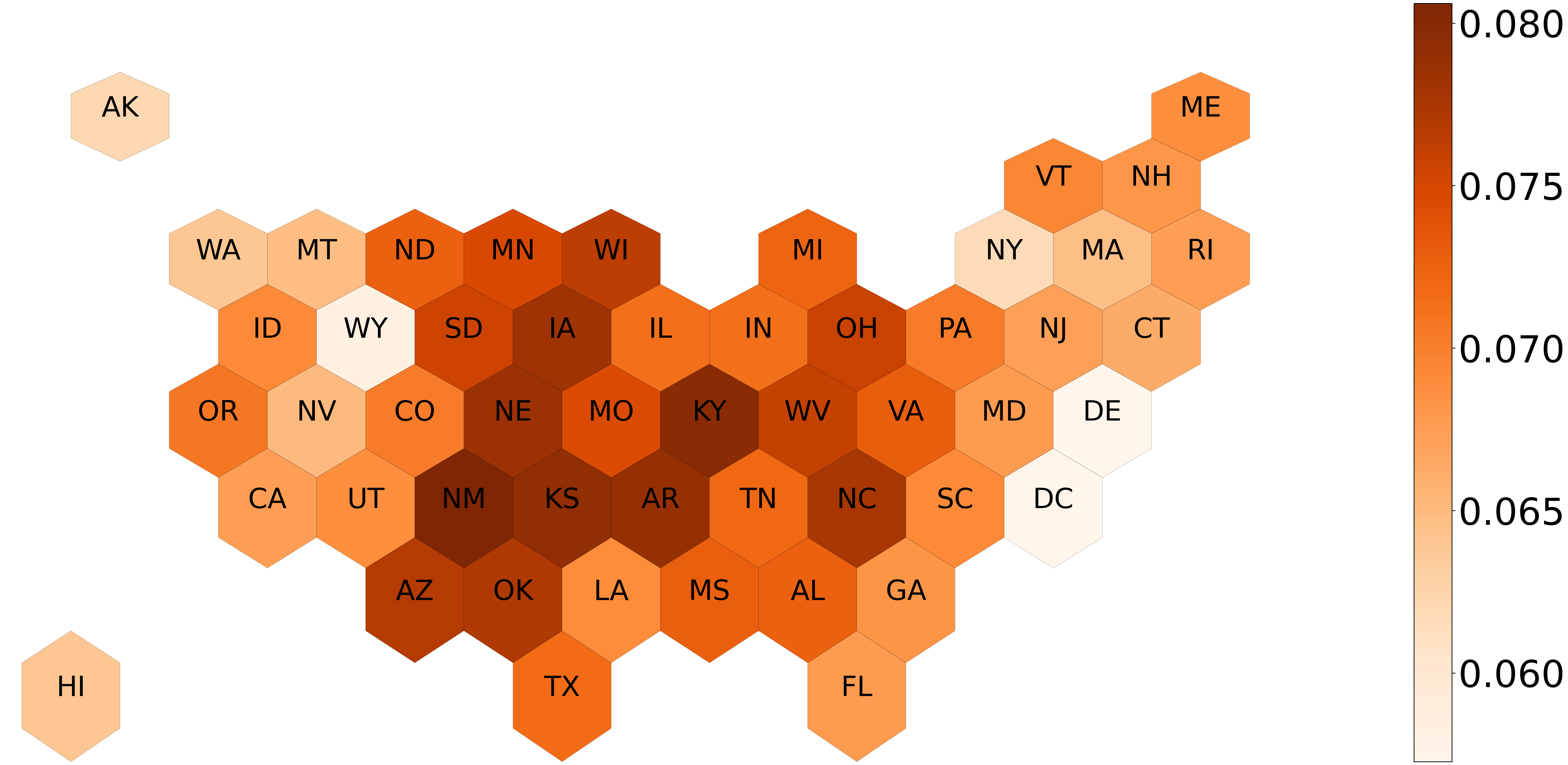}} 
    \subfigure[Pro-Science-Moderate]{\includegraphics[width=0.3\textwidth]{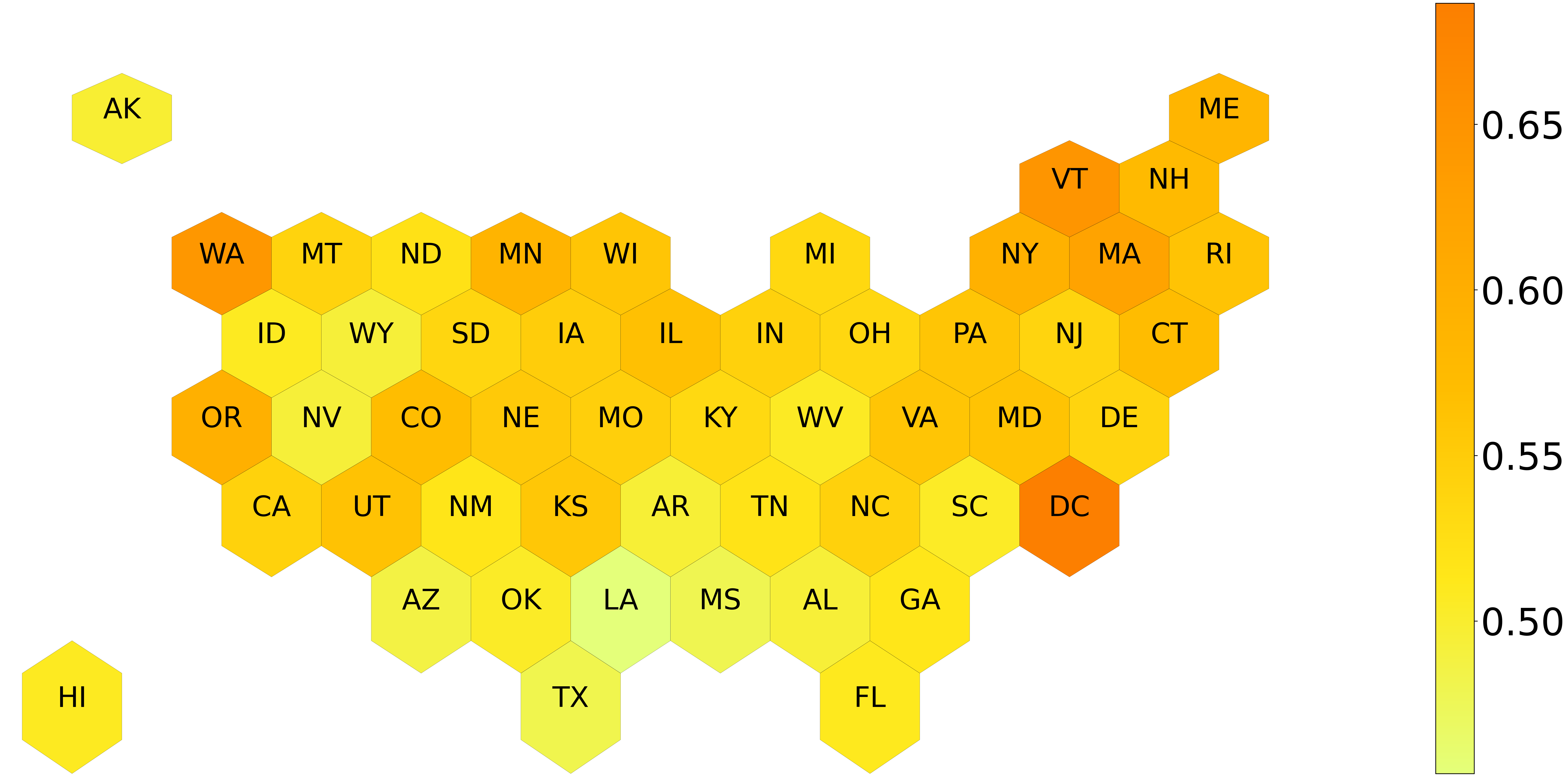}} 
    \subfigure[Pro-Science-Right]{\includegraphics[width=0.3\textwidth]{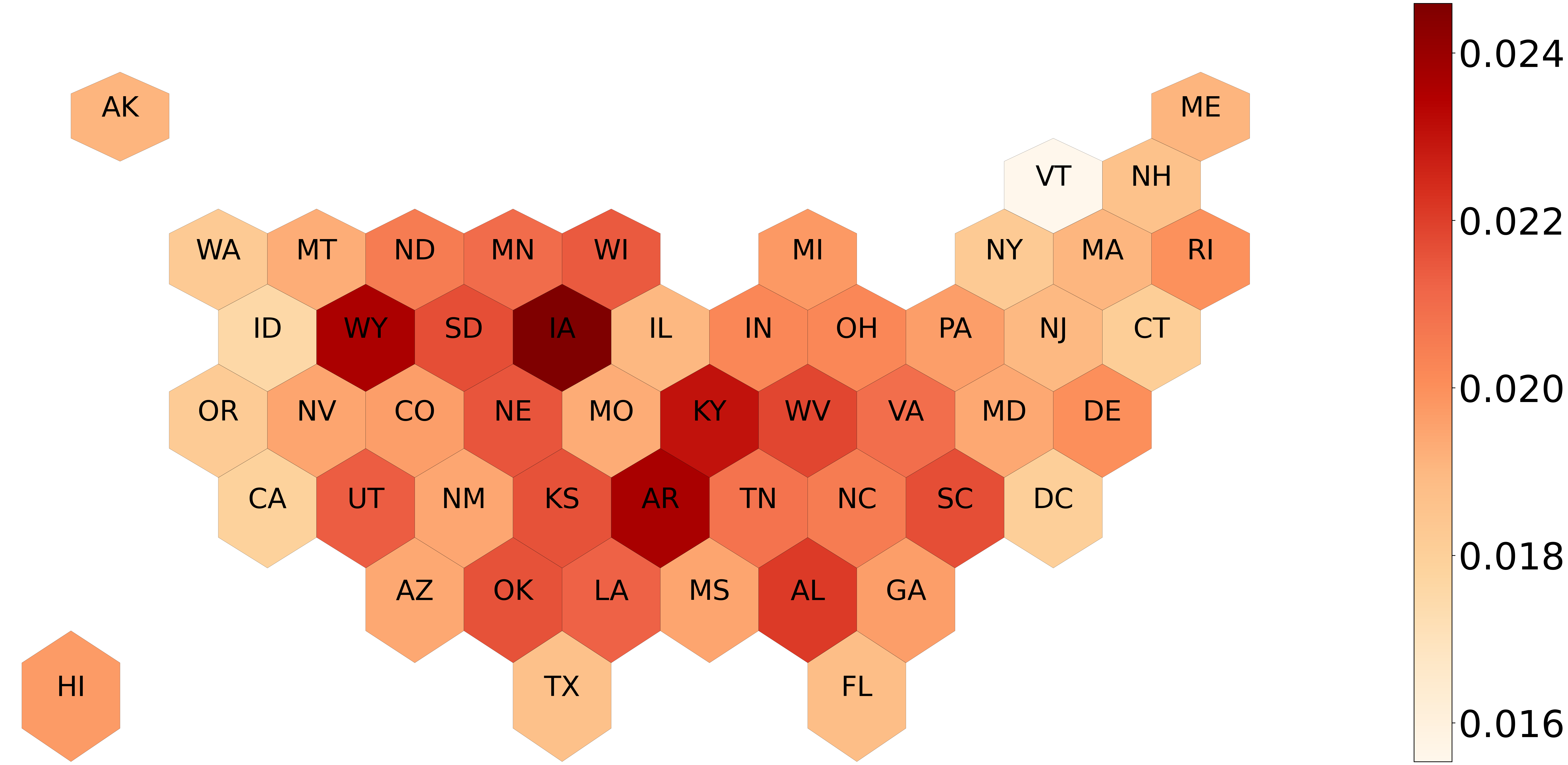}}
    \subfigure[Anti-Science-Left]{\includegraphics[width=0.3\textwidth]{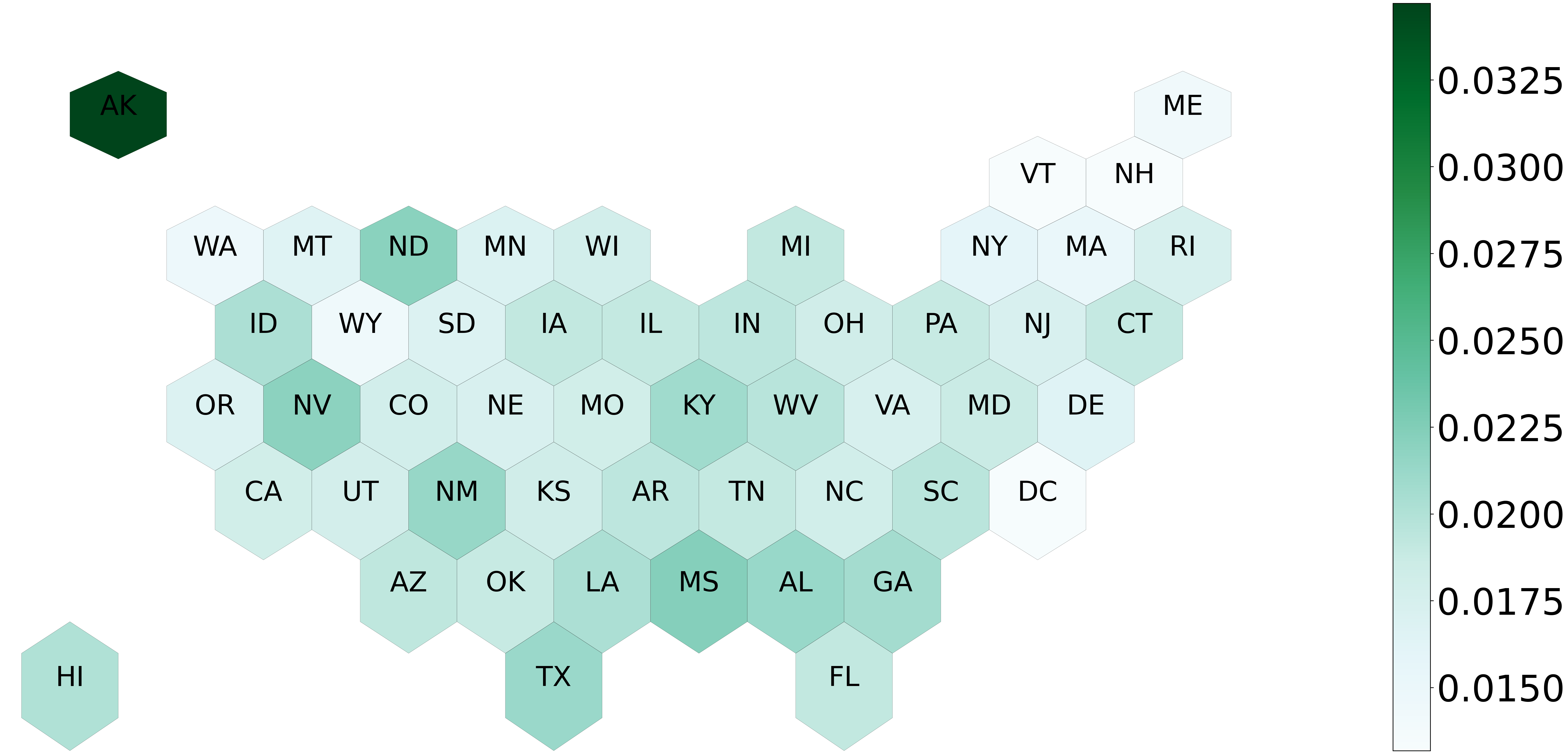}}
    \subfigure[Anti-Science-Moderate]{\includegraphics[width=0.3\textwidth]{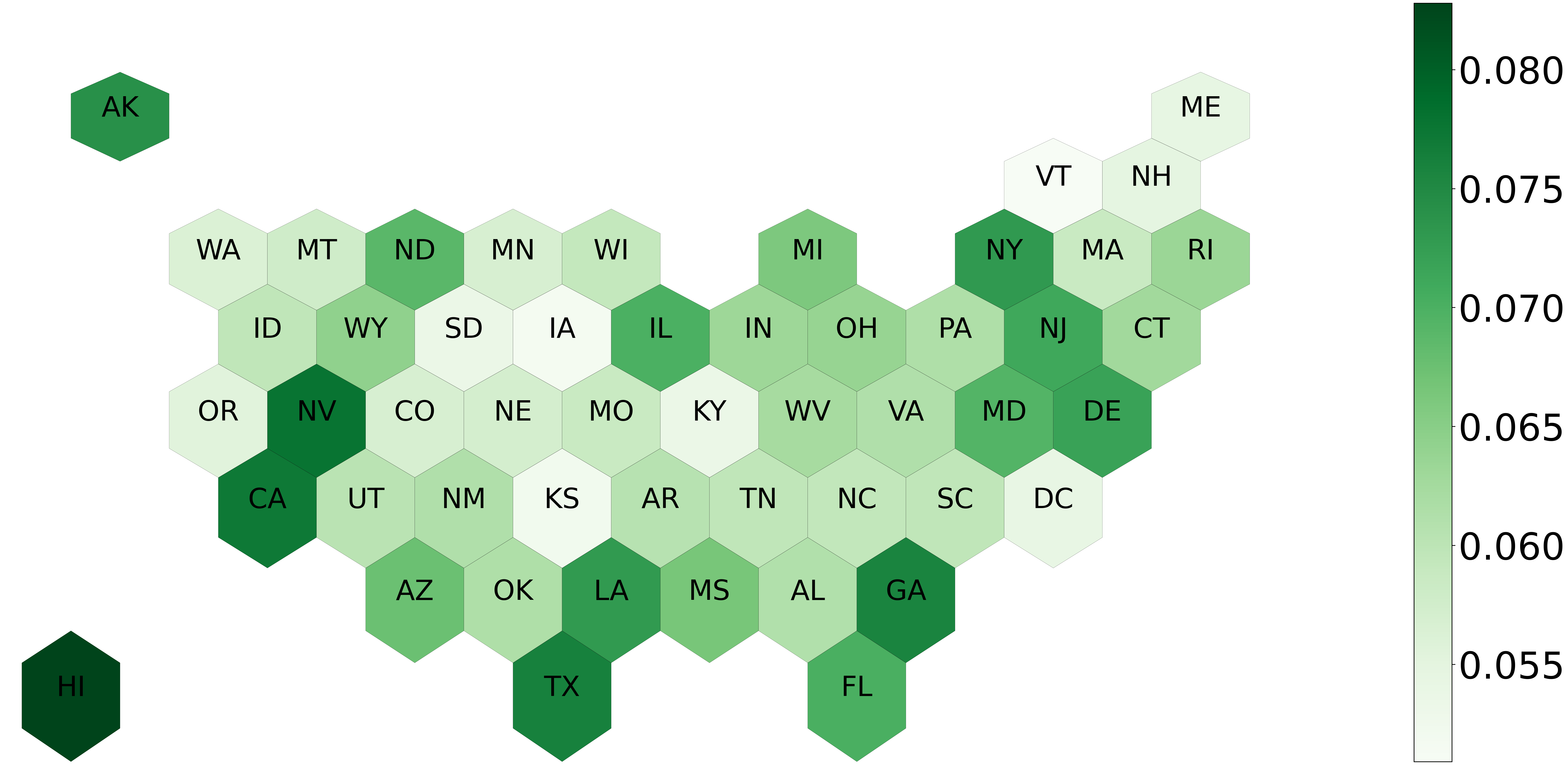}}
    \subfigure[Anti-Science-Right]{\includegraphics[width=0.3\textwidth]{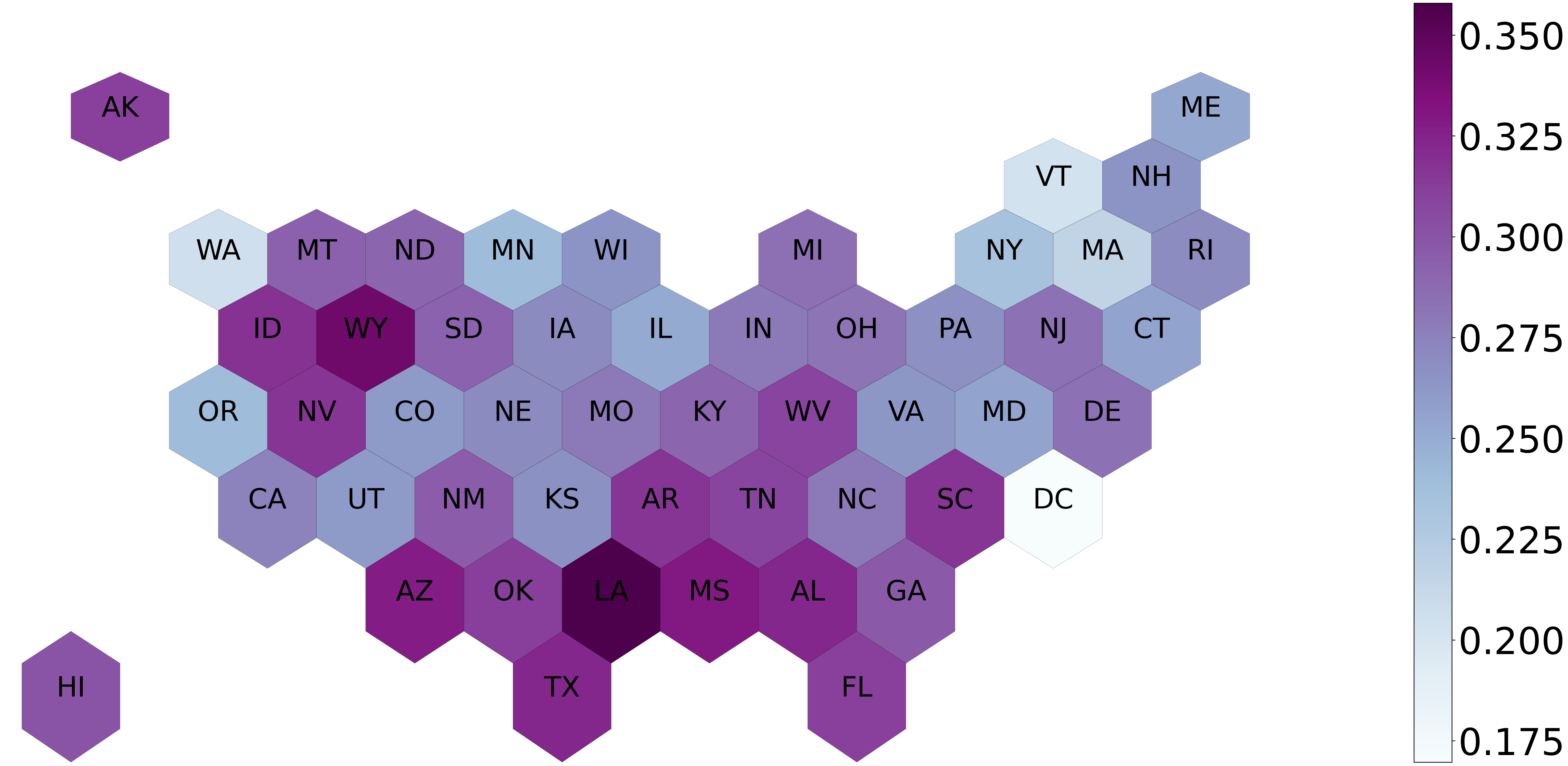}}
    \caption{Fraction of state's Twitter users per ideological category. Figures (a)-(c) show the fraction of states' Twitter users who are classified as Pro-Science Left, Pro-Science Moderate and Pro-Science Right, respectively. Figures (d)-(f) show the fraction of states' Twitter users who are classified as Anti-Science Left, Anti-Science Moderate and Anti-Science Right, respectively.
    }
    \label{fig:geodist}
\end{figure*}

\subsection{Dynamics of Polarization}



Research shows that opinions of Twitter users about controversial topics do not change over time~\cite{smith2013role}. To investigate whether user alignments along the three polarization dimensions change over time, we group tweets by time into seven biweekly intervals: {January 21--31,2020; February 1--15,2020; February 16--29, 2020; March 1--16, 2020; March 17--31, 2020; April 1--15, 2020; April 16-May 1, 2020}. 
There are 3.0K 
users who tweet consistently in all the $7$ biweekly intervals. For each of the $N$ users, we compute cumulative domain scores along Science, Political and Moderacy dimensions for all time intervals $t$ and  compute the average absolute change $\bar{\Delta}_{t,t-1}$ in domain score from biweekly period $t-1$ along each dimension given by, 

$$\bar{\Delta}_{t,t-1} = \frac{\Sigma_{i=1}^{N}|\delta_{i,t} - \delta_{i,t-1}|}{N}$$

where, $\delta_{i,t}$ represents the domain score for a user $i$ in biweekly period $t$. The small values of $\bar{\Delta}_{t,t-1}$ in Table \ref{tab:avg-abs-change} confirm that user alignments do not change significantly over time. 

\begin{table}
    \footnotesize
    \centering
    \begin{tabular}{ lllllll }
    \toprule
     \textbf{Dimension} & \textbf{$\bar{\Delta}_{2,1}$} & \textbf{$\bar{\Delta}_{3,2}$} &\textbf{$\bar{\Delta}_{4,3}$} &\textbf{$\bar{\Delta}_{5,4}$} &\textbf{$\bar{\Delta}_{6,5}$} &\textbf{$\bar{\Delta}_{7,6}$} \\
     \midrule
     Political & $0.09$ & $0.05$ & $0.03$ & $0.02$ & $0.03$ & $0.02$\\
     Science & $0.13$ & $0.07$ & $0.04$ & $0.02$ & $0.02$ & $0.02$\\
     Moderacy & $0.21$ & $0.11$ & $0.07$ & $0.04$ & $0.04$ & $0.03$\\
    \bottomrule
    \end{tabular}
    \caption{Average absolute change in domain score along consecutive biweekly intervals.}
    \label{tab:avg-abs-change}
\end{table}

Although individual's alignments do not change, the number of users within each ideological group does change over time. User alignments do not change, therefore we leverage polarization classification results to show biweekly fractions of active users per ideological category. 
Figure \ref{fig:pol-over-time} shows the composition of active users in all categories. As time progresses, we can clearly see the growth in the Pro-Science-Moderate category accompanied by a corresponding decline in Anti-Science-Right users.  

\subsection{Topics of Polarization}


To better understand what each of the six groups tweets about, we  collect the 50 most frequent hashtags used by each group, after removing hashtags common to all six groups. 
Figure \ref{fig:hashtag-wordcloud} shows the wordclouds of the most common hashtags within each group, sized by the frequency of their occurrence.
Most striking is the use of topics related to conspiracy theories, such as \textit{\#qanon}, \textit{\#wwg1wga}, by the Anti-Science-Right group, along with politically charged references to the \textit{\#ccpvirus} and \textit{\#chinavirus}. This group also uses hashtags related to President Trump's re-election campaign, showing the hyper-partisan  nature of COVID-19 discussions. Another partisan issue appears to be \textit{\#hydroxychloroquine}, a drug promoted by President Trump. It shows up  in both Pro-Science-Right and Anti-Science-Right groups, but is not discussed by other groups. 

The polarized nature of the discussions can be seen in the user of the hashtags \textit{\#trumppandemic} and \textit{\#trumpvirus} by the Left and Pro-Science groups. However, in contrast to Anti-Science groups, Pro-Science groups talk about COVID-19 mitigation strategies, using hashtags such as \textit{\#stayhomesavelives}, \textit{\#staysafe} and \textit{\#flattenthecurve}.

\subsection{Geography of Polarization}
Responses to the coronavirus pandemic in the US have varied greatly by state. While governors of New York, California, Ohio and Washington reacted early by ordering lockdowns, governors of Florida and Mississippi have downplayed the gravity of the situation for a longer time. 
To explore the geographical variation in ideological alignments, we group users by the state from which they tweet and compute the fraction of their respective state's Twitter users belonging to an ideological group. We then generate geo-plots, shown in Figure \ref{fig:geodist}, to highlight the ideological composition of each state. 

We see a higher composition of Pro-Science-Moderates (Figure \ref{fig:geodist}(b)) in Washington, Oregon, DC, Vermont. As expected, these states have a lower fraction of Anti-Science users as can be seen from Figures \ref{fig:geodist} (d),(e) and (f). Governors of these states were quick to enforce lockdowns and spread pandemic awareness amongst the general public. 

Over the course of the pandemic, we have seen the strong opposition to masking mandates and closing down of businesses in California, Nevada, Hawaii, Georgia and Texas. These anti-science sentiments are reflective in Figure \ref{fig:geodist} (e), which shows that these states have a comparatively higher proportion of their Twitter users in the Anti-Science-Moderate ideology group. 

States such as South Carolina, Mississippi, Louisiana, Wyoming, Texas and Arizona have had chequered responses to the pandemic with political and religious leaders consistently downplaying the pandemic. The geo-plot is reflective of the same with these states have a higher fraction of Anti-Science hardline-right users. 

\section{Conclusion}


Our analysis of a large corpus of online discussions about COVID-19 confirm the findings of opinion polls and surveys \cite{pew2020partisan}: opinions about COVID-19 are strongly polarized along partisan lines. Political polarization strongly interacts with attitudes toward science: conservatives are more likely to share conspiracy and anti-science information, while liberals and moderates are more likely to share information from pro-science sources. On the positive side, we find that the number of pro-science, politically moderate users dwarfs other ideological groups, especially anti-science groups. This is reassuring from the public health point of view, suggesting that a plurality of people are ready to accept scientific evidence and trust scientists to lead the way out of the pandemic. The geographical analysis of polarization identifies regions of the country, particularly in the South and the West where anti-science attitudes are more common. Messaging strategies should be tailored in these regions to communicate with science skeptics.

A larger issue of polarization is that it creates societal-scale vulnerabilities by amplifying distrust of authorities and making it easier for malicious actors to influence public opinion through concentrated digital misinformation campaigns~\cite{Thorp1405,Ferrara2020}. Amplifying the negative opinions about face coverings or vaccines within even a fraction of the population could create a cascade of adverse effects. Once polarization gets a life of its own it can lead to political instability, paralysis, and at worst democratic erosion \cite{conover2011political,mccoy2016polarized}. If societal trust is weakened, cynicism towards politics intensifies and confidence in public institutions wanes ~\cite{mccoy2016polarized}. Because we find a slow-but-steady increase in moderate users, we believe that ongoing efforts to inform the public could be benefiting user behavior online.

Although we show good performance on classifying polarized opinions, additional work is required to infer finer-grained opinions. Namely, by predicting fine-grain polarization among users, we could better infer, for example, network effects such as whether users prefer to interact with more polarized neighbors. Moreover, longer-term trends need to be explored in order to better understand how opinions change dynamically. This will better test whether whether social influence or selective formations of ties are the drivers of echo chambers and polarization. Finally, we want to explore polarization across countries to understand how different societies and governments are able to address polarization and how these polarized dimensions relate to one another across the world. 

\bibliographystyle{aaai}
\bibliography{Polarization}
\balance


\end{document}